# Permanent magnet system to guide superparamagnetic particles

Olga Baun and Peter Blümler*

Institute of Physics, University of Mainz, Germany

*corresponding author: email: bluemler@uni-mainz.de

**Abstract:**

A new concept of using permanent magnet systems for guiding superparamagnetic nano-particles on arbitrary trajectories over a large volume is proposed. The basic concept is to use one magnet system which provides a strong and homogeneous (dipolar) magnetic field to magnetize and orient the particles. A second constantly graded (quadrupolar) field is superimposed to the first to generate a force on the oriented particles. In this configuration the motion of the particles is driven solely by the component of the gradient field which is parallel to the direction of the homogeneous field. As a result particles are guided with constant force and in a single direction over the entire volume. The direction can simply adjusted by varying the angle between quadrupole and dipole. Since a single gradient is impossible due to Gauß' law, the other gradient component of the quadrupole determines the angular deviation of the force. However, the latter can be neglected if the homogeneous field is stronger than the local contribution of the quadrupole field. A possible realization of this idea is a coaxial arrangement of two Halbach cylinders. Firstly a dipole produces a strong and homogeneous field to evenly magnetize the particles in the entire volume. The second cylinder is a quadrupole to generate the force. The local force was calculated analytically for this particular geometry and the directional limits were analyzed and discussed. A simple prototype was constructed to demonstrate the principle on several nano-particles, which were moved along a rough square by manual adjustment of the force angle.

The observed velocities of superparamagnetic particles in this prototype were always several orders of magnitude higher than the theoretically expected value. This discrepancy is attributed to the observed formation of long particle chains as a result of their polarization by the homogeneous field. The magnetic moment of such a chain is then the combination of that of its constituents, while its hydrodynamic radius stays low.

A complete system will consist of another quadrupole (third cylinder) to additionally enable scaling of the gradient/force strength by another rotation. In this configuration the device could then also be used as a simple MRI machine to image the particles between movement intervals. Finally, a concept is proposed by which superparamagnetic particles can be guided in three-dimensional space.

# 1. Introduction

Moving (superpara-)magnetic nano-particles (SPP) remotely and contactless along arbitrarily chosen trajectories inside biological systems at reasonable speeds could assist many therapeutic or diagnostic or jointly theranostic methods [1, 2]. These minimal invasive applications are summarized in recent reviews: One major field is magnetic drug targeting (MDT) [3] where highly potent pharmaceuticals are concentrated at the disease site (tumor, infections, thrombi, etc.) simultaneously minimizing deleterious side effects. The prime example are SPPs with chemotherapeutics or genetic vectors as a payload in cancer therapy [4, 5]. Another interesting idea is to use local concentrations of SPPs to destroy tissue via local heating using strong AC-fields (aka "hyperthermia") [6]. Finally, magnetic forces are also used to separate magnetically labeled materials from mixtures either for diagnostics [7] or purification purposes [8, 9].

Another medical application of magnetic particles has to be mentioned, which is the use of mainly superparamagnetic iron oxide (SPIO) nanoparticles as contrast agents in clinical MRI (**m**agnetic **r**esonance **i**maging). This is a well-established methodology [10-12] utilizing the local disturbance of the magnetic field in the vicinity of SPPs with strong influences on the relaxation times of surrounding nuclear spins. The MRI signal then allows the detection of very low concentrations [13] and surface functionalization also enables molecular targeting [12].

In the last two decades great advances have been made in all these applications of SPPs due to the flourishing nanotechnology. However, most of the research in this field was dedicated to synthesis and biomedical modifications of the nanoparticles [1, 14-16], while until recently the development of magnetic steering methods and systems was more exclusive as for many demonstrations it was sufficient to apply simple bar magnets to test tubes. However, a magnet applied from only one side has the inherent problem that the magnetic field changes non-linearly over distance exerting a non-constant force on the particles. For applications on animals or even humans this approach is therefore very limited, because the forces increase towards the surface of the magnet and hence the surface of the organism. This approach makes controlled movements inside deeper tissues very problematic if not impossible. Additionally, a variation of directions can be difficult to achieve as access might be limited depending on the position on the body. Therefore, the main application was retaining otherwise free circu-lating nanoparticles in regions close to the skin by placing strong magnets in direct vicinity of the disease focus [3].

Again, this approach is limited to one direction over short distances, $r$, because a single magnet can only pull particles towards it with a field strength that spatially decays with $r^{-3}$, and hence its gradient/force drops with $r^{-4}$. To overcome these severe limitations clever designs of e.g. pushing magnets [3, 17] and optimized magnetization distribution [18] have been demonstrated. They are nicely reviewed in [3]; and in a similar review by Shapiro et al. [19] the current state of art is concluded as follows: „*Overall, one of the biggest open challenges in magnetic delivery is precisely targeting deep tissue targets – there are as yet no imaging and actuation systems that can achieve the external-magnet deep-focusing (…). To achieve deep targeting requires solution of, at least, four major issues: (1) sufficient magnetic fields/forces deep in the body, (2) real-time imaging, (3) sophisticated control algorithms, and (4) mathematical modeling of the carrier motion in vivo…*"

The question is: Can a magnet system for such a purpose be sketched and what general properties must it have? To really guide particles in more than one direction directly suggests the use of a surrounding magnet system [20-22], because single sided magnet systems do not have sufficient degrees of freedom to generate controlled fields inside the body, and the rapid field decay would also make these magnets either too weak or too bulky. With regards to the last (weight/size) argument the magnet system should probably also be made from permanent magnets. Permanent magnets have several properties which make them more practical than electromagnets for MDT applications. First of all their magnetic energy is permanently stored hence they do not need electrical power. To highlight this argument, one can estimate that the magnetic field produced by about 1 cm$^3$ of modern rare earth magnets (e.g. NdFeB) would require electric power in the order of some kW when produced by an electromagnet, ignoring the extra power for cooling. Other advantages might be price and comfort of use. However, their permanent magnetic field is also regarded as a disadvantage because it cannot be regulated or simply switched off. Obviously this is possible using electromagnets, however not as simple as it sounds, because of inductance and changes in resistance, geometry etc. due to heating effects introduce severe nonlinearities in real applications [21].

Whilst the argument of nonadjustable magnetic strength holds for single permanent magnets, they can be arranged to larger assemblies which then allow for scaling their magnetic field typically by mutual rotation [23-25]. Due to the extreme magnetic hardness of modern rare-earth permanent magnets the resulting fields are reproducible, easy to calculate theoretically or at least simply to calibrate. For all these reasons it should be most favorable in terms of strength per weight, power consumption and controllability to build magnetic guiding devices for MDT from permanent magnets.

The aim of this publication is to present a concept how a device made solely from permanent magnets could generate strong but homogeneous forces over large volumes, which can be precisely controlled in strength and direction. These forces are described by simple equations which in principle need no adjustments or time-dependent corrections. At worst the magnetic field of a real device, which of course will deviate from theory, needs to be measured once and taken into account. In addition, the proposed instrument can basically also be used for MRI by simply adding a tuned rf-coil. This would allow to take images of the particles at intervals during the moving process to control a successful operation in situ and in vivo. For these reasons the instrument was nicknamed "Mag-Guider" (**Mag**netic **Guid**e and Scann**er**).

To demonstrate the concept, a simplified prototype was constructed which clearly showed that SPPs can indeed be guided in a controlled manner. However, the full potential of all the presented ideas has not been exploited yet.

## 2. Concepts and Theory

### 2.1. Magnetic force

In order to guide a SPP at a spot, $\vec{r}$, it has to experience a magnetic force,

$$\vec{F}_{mag}(\vec{r}) = \vec{\nabla}\left(\vec{m}(\vec{r}) \cdot \vec{B}(\vec{r})\right). \qquad [1]$$

This is the gradient of the magnetic field, $\vec{B}$ [T], acting on a particle with a magnetic moment, $\vec{m}$ [Am$^2$]. The latter can be estimated from its volume integral, i.e. the bulk mass magnetization, $\vec{M}$ [Am$^2$/kg or emu/g],

$$\vec{m} = \rho V \vec{M} = \frac{\rho V \chi}{\mu_0} \vec{B} \quad \text{and} \quad \lim_{B \to \infty} \vec{m} = \rho V \vec{M}_{sat}, \qquad [2]$$

where $\rho$ is the density [kg/m$^3$] and $V$ the volume [m$^3$] of the particle. The dependence of the magnetization $M = M_R + M(B)$ on the applied magnetic field can be complex and generally has a hysteresis. Typically $M(B)$ is approximated by the monotonic Langevin-function and the remnant part, $M_R$, is ideally zero or very close to zero for SPP and saturates, $M_{sat}$, at high fields. Alternatively, the magnetic susceptibility, $\chi$ [dimensionless], can be used to character-ize SPPs by their bulk properties. Furthermore, the nature of SPPs as small more or less isolated particles will cause an alignment of themselves or of their magnetic moment with the local magnetic field, hence

$$\vec{m} \parallel \vec{B} \quad \text{or} \quad \vec{m} \cdot \vec{B} = |\vec{m}||\vec{B}|. \quad [3]$$

Therefore, eq. [1] can be simplified to

$$\vec{F}_{\text{mag}}(\vec{r}) = |\vec{m}(\vec{r})| \, \vec{\nabla} |\vec{B}(\vec{r})|, \quad [4]$$

because the spatial variation of the magnetic moment inside a particle does not contribute to the force or $\vec{\nabla}|\vec{m}| \approx 0$. The fact that the force depends on the gradient of the norm of $\vec{B}$ explains why the particles move from low to high field intensities (independent of the field polarity) [3].

It is noteworthy to give realistic estimates of $F_{\text{mag}}$ for SPIOs which are typically in the range from $10^{-25}$ to $10^{-11}$ N [3] and hence it is advisable to maximize $|\vec{m}|$ as well as $\vec{\nabla}|\vec{B}|$.

## 2.2. Conceptual magnet design

Equation [4] is still spatially dependent which can complicate the control of motion tremendously [21] due to the generally non-linear dependence of magnetic fields over space. In order to simplify this scenario, one would like to have a strong and homogeneous magnetic field, $\vec{B}(\vec{r}) = \vec{B}_0$, to maximize $|\vec{m}(\vec{r})| = \rho V M_{\text{sat}}$ and evenly orient the particles. A homogeneous field would of course result in $\vec{\nabla}|\vec{B}| = 0$ and thus $F_{\text{mag}} = 0$. Since magnetic fields can be added another graded field can be superimposed, ideally with $\vec{F}_{mag}(\vec{r}) \propto \vec{\nabla}|\vec{B}(\vec{r})| = \vec{G}$, i.e. causing a constant force over the sample's volume.

However, this combination does more than that. Within certain limits (see 2.5) it creates a spatially constant force along a single direction over a larger volume, although Gauß' law, $\vec{\nabla} \cdot \vec{B} = 0$, forbids a single gradient component. Nonetheless, the combination of a strong homogenous with a constantly graded magnetic field leaves only that gradient component which is parallel to $B_0$ to be effective. This can be understood from eq. [3] where $\vec{m}$ orients along the direction of $\vec{B}(\vec{r})$ which is dominated by $\vec{B}_0$. Then the scalar product in eq. [1] will select essentially only that component of the graded field which is parallel to $\vec{B}_0$. As a result, the particles can be moved in a single, well defined and adjustable direction, which is essentially spatially independent.

This is a situation well known in MRI where the much stronger and homogeneous polarization field allows the selection of a single, parallel component from the spatially encoding fields produced by the gradient coils. On the contrary, the failure of this concept (e.g. at low polarization fields) is typically denoted as "gradient tensor imaging" due to concomitant gradients of similar strength [26].

## 2.3. Realization with Halbach di- and quadrupole to select force direction

One possibility to realize a homogeneous and a constantly graded magnetic field is the combination of Halbach cylinders of different polarity. This is a cylinder where at every azimuthal angle θ the magnetization direction of the permanent magnetic material of the cylinder wall changes by an angle $(1 + p/2)\,\theta$ where $p$ is the polarity, e.g. $p = 2$ for a dipole and $p = 4$ for a quadrupole [27, 28]. In order to generate a homogeneous field a Halbach dipole (superscript "D") is needed, as shown in Fig. 1a. The resulting homogeneous field is then given by

$$\vec{B}^D(\vec{r}) = B_0 \begin{bmatrix} 0 \\ 1 \end{bmatrix} \quad \text{with} \quad \vec{r} = \begin{bmatrix} x \\ y \end{bmatrix} \quad \text{and} \quad B_0 = B_R^D \ln \frac{r_i^D}{r_o^D} \qquad [5]$$

if its direction is chosen along the $y$-direction. Here $B_R$ is the remanence of the permanent magnetic material and $r_i$ is the inner and $r_o$ the outer radius of the cylinder.

On the other hand a graded field is produced by a quadrupole (superscript "Q")

$$\vec{B}^Q(\vec{r}) = \mathsf{G} \cdot \vec{r} = G \begin{bmatrix} -1 & 0 \\ 0 & 1 \end{bmatrix} \vec{r}, \quad \text{with} \quad G = 2 B_R^Q \left( \frac{1}{r_i^Q} - \frac{1}{r_o^Q} \right). \qquad [6]$$

The norm of its field has the form of a cone (see Fig. 1b), which can be decomposed in two spatially linear components (in $x$ and $y$ direction, see Fig. 1c and 1d) both having the same derivative $G$.

Dipole and quadrupole cylinder can then be mounted such that they surround one another coaxially allowing free rotation. If then the quadrupole is rotated by an angle α relative to the dipole the resulting field is given by superposition, i.e. the components of the gradient field rotate by 2α (derivation see [25]).

$$\vec{B}(\vec{r}) = B_0 \begin{bmatrix} 0 \\ 1 \end{bmatrix} + G \begin{bmatrix} -x\cos 2\alpha + y\sin 2\alpha \\ x\sin 2\alpha + y\cos 2\alpha \end{bmatrix} \qquad [7]$$

If further $B_0 > G\,r$, with $r$ as the particle position, only the $y$-component (direction of $B_0$) acts as a force on the particle which is then moving along a direction given by 2α and the equation is independent of $r$. This angle can be freely selected by mechanical rotation of the quadru-pole relative to the dipole, and since Halbach cylinders have no stray field, this rotation is ideally force-free[23, 24, 29].

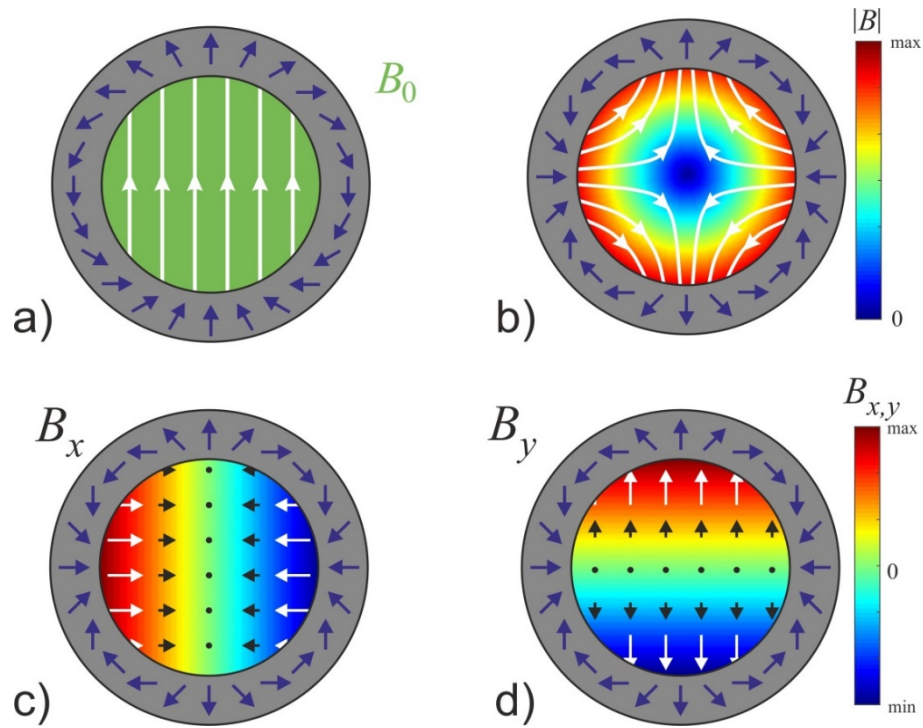

Fig. 1: Schematic illustration of the Halbach magnet geometries: a) Inner Halbach dipole field created by an annular cylinder of permanent magnet material (dark gray) with continuously changing magnetization direction as indicated by the blue arrows. The rainbow colored shade in the center illustrates the amplitude of the field, while its direction is specified by the white arrowed flux lines. Since the field is homogeneous in a it has only a single (green) color. b) Inner Halbach quadrupole in the same representation as a). c) and d) show the $B_x$ and $B_y$ component of the flux in b). Here the black/white arrows indicate strength and direction of the flux (no flux lines!). Note the different colorbars in a/b) and c/d).

## 2.4 A second quadrupole for force scaling

To scale the force generated by the quadrupole its field can be (partially) cancelled by superposition of a second Halbach quadrupole (subscripts "Q1" and "Q2") which can also be freely rotated coaxially. For an angle β between both quadrupoles the field becomes (see Fig. 2a) [25]

$$\bar{B}(\bar{r}) = B_0 \begin{bmatrix} 0 \\ 1 \end{bmatrix} + G_{Q1} \begin{bmatrix} -x\cos 2\alpha + y\sin 2\alpha \\ x\sin 2\alpha + y\cos 2\alpha \end{bmatrix} + G_{Q2} \begin{bmatrix} -x\cos 2(\alpha+\beta) + y\sin 2(\alpha+\beta) \\ x\sin 2(\alpha+\beta) + y\cos 2(\alpha+\beta) \end{bmatrix}. \quad [8]$$

If then the radii and remanences of both quadrupoles are chosen such that $G_{Q1} = G_{Q2} = G$ equation [8] has two extremes. For β = 0 both quadrupoles are parallel and produce twice the gradient, $2G$, at an angle $2\alpha$ relative to the dipole (see Fig. 2b), while they cancel completely[1] for $\beta = 90°$ (see Fig. 2c). Then the remaining field is only that of the dipole. Therefore, the angle $\beta$ can be used to scale the magnitude of the gradient from 0 to $2G$ according to

$$\left|\vec{G}\right| = G\sqrt{2 + 2\cos 2\beta} = 2G\left|\cos\beta\right| \qquad [9]$$

Substituting eq. [4] and [8] in eq. [1] gives then the general equation for the force in the cylinder plane $(x,y)$

$$\vec{F}_{mag}(x,y) = \frac{\rho V\,M(B(x,y))}{\Xi}\begin{bmatrix}\left(G_{Q1}^2 + 2G_{Q1}G_{Q2}\cos 2\beta + G_{Q2}^2\right)x + B_0 G_{Q1}\sin 2\alpha + B_0 G_{Q2}\sin(2\alpha + 2\beta)\\ \left(G_{Q1}^2 + 2G_{Q1}G_{Q2}\cos 2\beta + G_{Q2}^2\right)y + B_0 G_{Q1}\cos 2\alpha + B_0 G_{Q2}\cos(2\alpha + 2\beta)\end{bmatrix}$$

with $\quad \Xi \equiv \Big[B_0^2 + \left(G_{Q1}^2 + 2G_{Q1}G_{Q2}\cos 2\beta + G_{Q2}^2\right)(x^2 + y^2) +$

$$+\,2B_0\left(G_{Q1}x\sin 2\alpha + G_{Q1}y\cos 2\alpha + G_{Q2}x\sin(2\alpha + 2\beta) + G_{Q2}y\cos(2\alpha + 2\beta)\right)\Big]^{\frac{1}{2}} \qquad [10]$$

which contains only technical constants with the exception of $M(B(x,y))$. If however $B_0$ is chosen such that it exceeds or is close to the saturating field the field dependence of the magnetization also vanishes and becomes a constant $M_{sat}$. If additionally $G_{Q1} = G_{Q2} = G$ and $\beta = 0$ eq. [10] simplifies to

$$\vec{F}_{mag}(x,y) = \frac{2\rho V M_{sat} G}{\sqrt{B_0^2 + 4G^2(x^2 + y^2) + 4B_0 G(x\sin 2\alpha + y\cos 2\alpha)}}\begin{bmatrix}2Gx + B_0\sin 2\alpha\\ 2Gy + B_0\cos 2\alpha\end{bmatrix} \qquad [11]$$

and $\qquad \left|\vec{F}_{mag}\right| = \rho V M_{sat}\,2G \qquad [12]$

The last equation shows that the force is constant over the entire sample, i.e. independent of the position of the particle. A similar design of three nested dipoles, of which two were rotated to computer controlled positions, has already been realized for EPR-spectroscopy [24].

In equilibrium this magnetic force has to overcome the force of Stoke's friction, $F_{frc}$ in the viscous solution [30] and

$$\vec{F}_{mag} = \vec{F}_{frc} = 6\pi\eta R\vec{v} \qquad [13]$$

where $\eta$ is the dynamic viscosity [Pa s] of the liquid medium, $R$ is the hydrodynamic radius and $\vec{v}$ is the resulting velocity of the particle. To estimate the speed for the simplified situation of $G_1 = G_2 = G$ and $\alpha = \beta = 0$ the velocity of a spherical particle is given by

$$\bar{v} = \frac{4\rho R^2 M_{sat} G}{9\eta \sqrt{B_0^2 + 4G^2(x^2+y^2) + 4B_0 Gy}} \begin{bmatrix} 2Gx \\ 2Gy + B_0 \end{bmatrix} \qquad [14]$$

---

[1] This is because $\sin(2\alpha + 2\beta) = \sin 2\alpha \cos 2\beta + \cos 2\alpha \sin 2\beta$ and hence for $\beta = \pi/2$ $\sin(2\alpha + \pi) = -\sin 2\alpha + 0$. Analogously $\cos(2\alpha + \pi) = -\cos 2\alpha$.

The strongest dependence is therefore on the size (*R*) of the particle which enters eq. [14] quadratically while all other parameters behave more or less linearly.

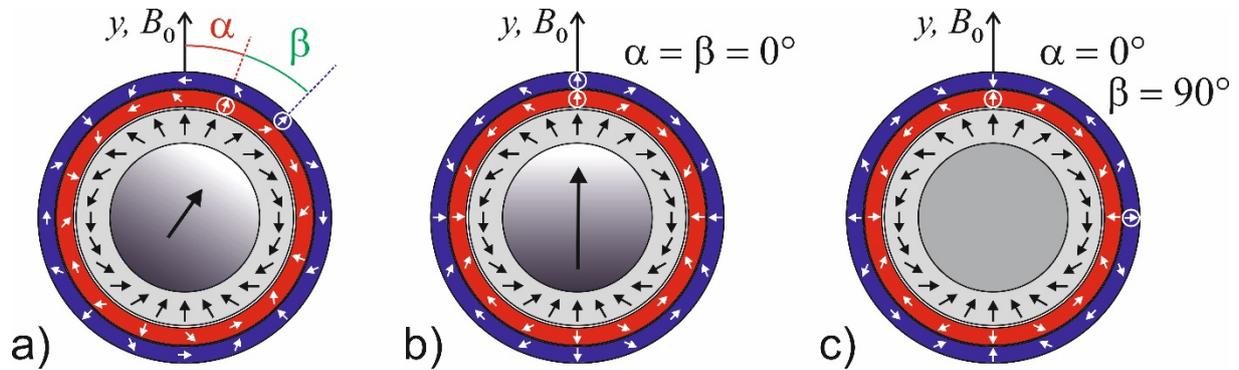

Fig. 2:  Schematic drawing of a dipole (gray central) surrounded by two quadrupoles (red and blue). a) The angle between the inner most quadrupole (red) and dipole is named α and the angle between both quadrupoles β. For better orientation the magnetization direction at the top of each quadrupole is encircled. b) The gradient becomes maximal for parallel aligned quadrupoles or β=0° and c) becomes minimal or even vanishes for orthogonal quadrupoles at β=90°.

## 2.5   Force direction and concomitant gradients:

The described system of one dipolar and two rotatable quadrupolar Halbach cylinders will generate a constant force with adjustable direction and strength over a larger area, but only if certain limits are met. As stated above a quadrupole generates two orthogonal gradient components which in the absence of a superimposed strong homogeneous field would only cause a homogeneous radial gradient without any preferential direction (see Fig. 1 b). If however the homogeneous field is always stronger than the local field-contribution of the gradient, its direction will dominate the concomitant other gradient. Unfortunately, the strength of this field contribution from the gradient grows proportional to the radius. Therefore, for a given strength of the homogeneous field there is a trade-off between the maximal gradient strength and the resulting angular precision, especially at high distances normal to the field direction of the homogeneous field, i.e. lateral rim of the sample.

A magnetic guiding system for MDT will need to generate gradients as strong as possible to maximize the force on small biocompatible SPP. Hence, the assumption that $B_0 \gg G\,r$ will be challenged (see Fig. 3).

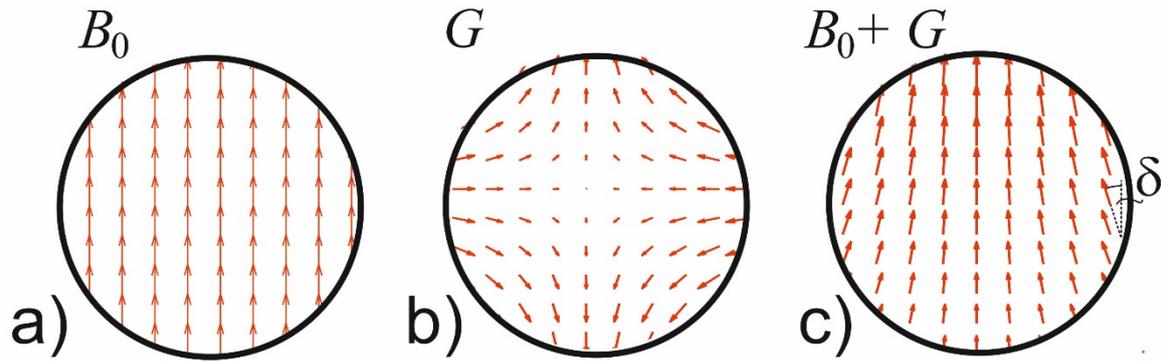

Fig. 3:  Schematic drawing to illustrate the concept of concomitant gradients. The resulting field is shown in c) which is the sum of a homogeneous dipolar field in a) and a quadrupolar gradient field in b). The red arrows point in the direction of the local magnetic flux while their lengths represents its strength. The deviation angle, δ, is shown on the right side of c, where the deviation is maximal.

As already stated this is a situation well known in (low field) MRI for which Yablonskiy et al. [31] calculated a critical radius,

$$R_c = \frac{B_0}{G}, \qquad [15]$$

at which the concomitant gradients, $G$, will bent spatially encoded structures in MRI-images. For typical MRI-conditions $R_c$ is much larger than the object size, so that the concomitant gradient can be neglected. Of course this is no longer possible for small $B_0$ or large $G$.

For simplicity's sake eq. [11] is used to search for the maximal angle at which the local force will deviate from the anticipated direction (all angles are defined relative to $B_0$, here the y-axis or α = 0). Since this equation was derived for two quadrupoles of equal strength, their total gradient is $2G$ and hence $R_c = B_0/(2G)$. Space shall be probed on circles with $[x, y] = r_p [\sin\varphi, \cos\varphi]$. Then eq. [11] can be rewritten as

$$\vec{F}_{mag}(\zeta,\varphi) = \begin{bmatrix} \sin\varphi + \zeta\cos(2\alpha) \\ \cos\varphi + \zeta\sin(2\alpha) \end{bmatrix} \qquad \text{with} \qquad \zeta \equiv \frac{R_c}{r_p} \qquad [16]$$

The rotation of the quadrupoles by an angle α will always cause $\vec{F}_{mag}$ to point along 2α in the center or origin, where there is no concomitant field (or try eq. [11] with $x = y = 0$). This direction or angle will be treated as the angular reference $\hat{e}_F \equiv [\sin(2\alpha), \cos(2\alpha)]$. The next step is to calculate the desired deviation angle, δ, of the local force $\vec{F}_{mag}(\zeta,\varphi)$ to this reference direction.

$$\cos\delta = \frac{\vec{F}_{\text{mag}}(\zeta,\varphi)\cdot\hat{e}_F}{\left|\vec{F}_{\text{mag}}(\zeta,\varphi)\right|} = \frac{\zeta+\cos(2\alpha-\varphi)}{\sqrt{1+\zeta^2+2\cos(2\alpha-\varphi)}} = \frac{\zeta+\cos\phi}{\sqrt{1+\zeta^2+2\cos\phi}} \qquad [17]$$

The question is now to find the position, $\phi = 2\alpha-\varphi$, at which this deviation becomes maximal

$$\frac{\partial\cos\delta}{\partial\phi} = -\frac{\sin\phi(1+\zeta\cos\phi)}{\left(1+\zeta^2+2\zeta\cos\phi\right)^{\frac{3}{2}}} = 0 \qquad [18]$$

Maxima at $\phi_{max} = 0, \pi, 2\pi, \ldots$ are found, which is expected because these are angles $\varphi = 2\alpha - \phi_{max}$ at the central line of the rotated quadrupole, here $\cos\delta = 1$ or $\delta = 0$ and hence no deviation from $\hat{e}_F$ are observed. A minimum is found at $\cos\phi_{min} = -1/\zeta$ giving a maximal deviation by insertion into eq. [17] of

$$\cos\delta_{max} = \frac{\zeta-1/\zeta}{\sqrt{1+\zeta^2-2}} = \frac{1}{\zeta}\sqrt{\zeta^2-1} \qquad [19]$$

The maximal deviation angle is therefore (for $\zeta \geq 1$, or $R_c \geq r_p$)

$$\delta_{max} = \cos^{-1}\sqrt{1-\frac{r_p^2}{R_c^2}} = \cos^{-1}\sqrt{1-\frac{r_p^2 G^2}{B_0^2}} \quad \text{or}$$
$$G = \frac{B_0}{r_p}\sqrt{1-\cos^2\delta_{max}} \qquad [20]$$

This equation clarifies that absolute angular precision, $\delta_{max} = 0$, implies $G = 0$ and hence no forces. Any gradient will generate a more or less strong deviation from the force angle in the center particularly orthogonal to this direction ( $\varphi = 2\alpha - \pi/2 + \cos^{-1}(r_p/R_c)$ ) and towards the rim of the sample. As it can be seen in the discussion above ($\varphi = 2\alpha - \phi_{max}$ ) and by inspection of Fig. 3c the angular deviation is much smaller towards the center of the reference force. Therefore, a strategy to bypass this problem could be to hold the particles in this central strip and direct them by a combined rotation of dipole and quadrupole ($\alpha = 0$) while rotation of the second quadrupole still allows for force scaling. This would enable the use of much stronger gradients without losing the directional precision, however somewhat for the sacrifice of simplicity and conceptual beauty.

In order to plug some realistic numbers in these equations two different designs are discussed. The first is a smaller benchtop version and the other a whole (human) body device. For both $Nd_2Fe_{14}B$ permanent magnets with a $B_R = 1.45$ T ($\rho = 7.5$ g/cm$^3$) are used:

1) <u>benchtop Halbach magnet</u> consisting of a dipole with inner diameter of 10 cm ($r_i^D = 0.05$ m), outer diameter of 16 cm ($r_o^D = 0.08$ m) and a height of 10 cm. This will roughly give $B_0^D$ = 0.5 T and a mass of permanent magnet material of ca. 9 kg (including supports etc. make 30 kg a more realistic value [32]). For a maximal deviation angle $\delta_{max}$ = 1° over the entire $r_i^D$ = 5 cm eq. [20] will then give a maximal gradient of $G$ = 0.17 T/m. Relaxing these constraints to $\delta_{max}$ = 5° already gives $G$ = 0.87 T/m and for $G$ = 1.74 T/m a $\delta_{max}$ = 10° has to be tolerated.

2) <u>whole body magnet</u> with $r_i^D$ = 0.3 m, $r_o^D$ = 0.4 m and a length of 0.5 m will give a $B_0^D$ = 0.3 T (magnet mass ca. 825 kg). The limiting gradient strength for $\delta_{max}$ = 1° is $G$ = 0.018 T/m, $\delta_{max}$ = 5° gives $G$ = 0.087 T/m and $\delta_{max}$ = 10° finally $G$ = 0.17 T/m.

This very crude estimate already shows that for large volumes and strong forces/gradients very heavy instrumentation has to be built and the masses above do not include the quadrupoles, supports, gears and motors. One possibility to reduce these masses is to partially abandon the angular precision, which allows reducing $B_0$ in the presence of strong forces.

## 2.6  Magnetic Resonance Imaging

The use of a strong and homogeneous magnetic field while applying field gradients which are variable in amplitude and direction, suggests exploiting this apparatus also as an MRI-scanner (The theory of MRI is explained elsewhere [10, 33]) to locate the position of the particles when desired. Especially with regards to the fact that SPPs are commonly used as MRI contrast agents and that a control of their position after an attempt of shifting them inside a body is of paramount importance for a successful MDT procedure.

Since the sensitivity of NMR/MRI-experiments scale with $B_0^{7/4}$ [33] a strong homogeneous magnetic field is even more important for this type of operation, but a field strength in a range of 0.1 to 1 T is not untypical for MRI. The nuclear spins in the sample, typically the protons in hydrogen, can be then be excited by an orthogonal magnetic field oscillating with a frequency given by

$$\omega = \gamma B_0 \qquad [21]$$

with $\gamma$ as the gyromagnetic ratio (for $^1$H: $\gamma$ = 42.576 MHz/T). This would correspond to a radio-frequency of $\omega$ = 4 - 42 MHz for the $B_0$ range specified above. This rf-frequency is usually transmitted and received by the same resonant coil, which –besides a NMR-spectrometer– is the only thing necessary to add to the suggested design of Halbach quadrupoles and dipole to generate 2D-images. This is because, the spatial information in MRI is created via a spatially varying magnetic field (i.e. in the simplest case a constant gradient) which has to be super-imposed to the homogeneous magnetic field. A Fourier-transformation of the acquired time signal in the presence of such a constant gradient field will then correspond to a projection of the sample along the direction of the gradient. If then the gradient direction is rotated (at least by 180°) in subsequent experiments, a two-dimensional image can be reconstructed (using a Radon-transformation, aka "filtered backprojection" [10]). Similar setups were already realized and discussed [25, 34].

Given a maximal sampling bandwidth for data acquisition of $\Delta\omega$, will then limit the imaging gradient according to

$$G_\text{I} = \frac{\Delta\omega}{\gamma\,\Delta r} \qquad [18]$$

For $\Delta\omega = 250$ kHz, $\Delta r = 2r_\text{i}^\text{D}$ this gives $G_\text{I} = 0.06$ T/m for the benchtop and $G_\text{I} = 0.01$ T/m for the whole body magnet from the previous section. These values are at least an order of magnitude smaller than the maximal gradients achievable for SPP moving and will probably not cause a big relocation of the particles during a full rotation of the gradients for imaging. However, this needs to be verified.

## 2.7 Three-dimensional control

The extension of the presented magnet design to three dimensions can be realized in two ways. Firstly the Halbach cylinders can be replaced by Halbach spheres [35] (having the same magnetization distribution like a cross-section through a Halbach cylinder but then rotated around the axis through two opposite poles to form a sphere). Although the magnetic field in such a hollow sphere will be stronger and possibly more homogeneous, they have the big disadvantage that the construction is very difficult and access to the interior is problematic. The latter can in solved by access holes or planes [35] or finding angles at which the sphere can be opened without force [28]. All these ideas are conceptually interesting because such an apparatus could generate magnetic forces at any angle on a static sample but is probably too difficult to realize.

On the other hand the second concept is rather simple. It should be sufficient to shorten the ideal (infinitely long) Halbach cylinder to create a field maximum in the third dimension which will attract the SPPs. Moving the particles then in this third dimension ($z$) could be simply done by placing the sample on a sliding table. Care has to be taken that the truncation of the Halbach cylinder does not destroy the field homogeneity in the cylinder plane ($x,y$) which can be prevented by adjusting the shape, size or remanence of the magnetic parts it is constructed from. Finally such a cylinder with a distinct maximum in the third dimension would also be advantageous for operating this system as a MRI-scanner, because the limited bandwidth of the resonant rf-field would only excite a small range of the magnetic field hence a thin slice in the third dimension. Therefore, this concept will probably not only reduce the mass of the system, but allow tomographic MRI of body regions where the SPPs are automatically attracted to. However, these ideas need to be demonstrated first.

## 3. Experimental

So far only ideal Halbach magnets were discussed. However the continuous change of the magnetization cannot be realized easily. Therefore, the usual implementation is a discretization into slabs with one magnetization direction. Details on discretization and construction of Halbach dipoles [28, 36, 37] and quadrupoles [25] can be found elsewhere.

As a proof of principle a first demonstration model for was constructed from Ni-coated NdFeB permanent magnets with a regular octagonal cross-section with a side length of 13.0 mm (inner diameter = 31.4 mm, outer diameter = 34.0 mm), a height of 19.5 mm and a remanence of $B_R$ = 1.398 T (grade 46MGOe from AR.ON GmbH in Mülheim a.d.R., Germany). They were not particularly chosen for this task but are left-overs from another project. Limited by a total number of about 50 of such magnets, it was decided to construct a Halbach dipole made from two cylindrical rings each constructed from 16 of these magnets positioned with their centers on a circle with radius of 84.7 mm. The two rings were separated by a gap of 41 mm to homogenize the central field. The magnets were glued into 5 mm deep sockets milled in aluminium support rings with an inner diameter of 130 mm and an outer diameter of 227 mm (see Fig. 4a and 4b). A quadrupole was designed using 8 of the same magnets clamped into another aluminium ring so that their centers are at a radius of 130.7 mm and fixed by brass stirrups (see Fig. 4a and 4c). This quadrupole ring was mounted between the two layers of the dipole on POM-spools such that it can freely rotate around the dipole. No forces were noticed when rotating this ring by hand. The entire assembly had a mass of 9.5 kg (see Fig. 4a, d, e).

A second quadrupole for scaling the forces was not installed, because it would require more magnets of preferably larger size.

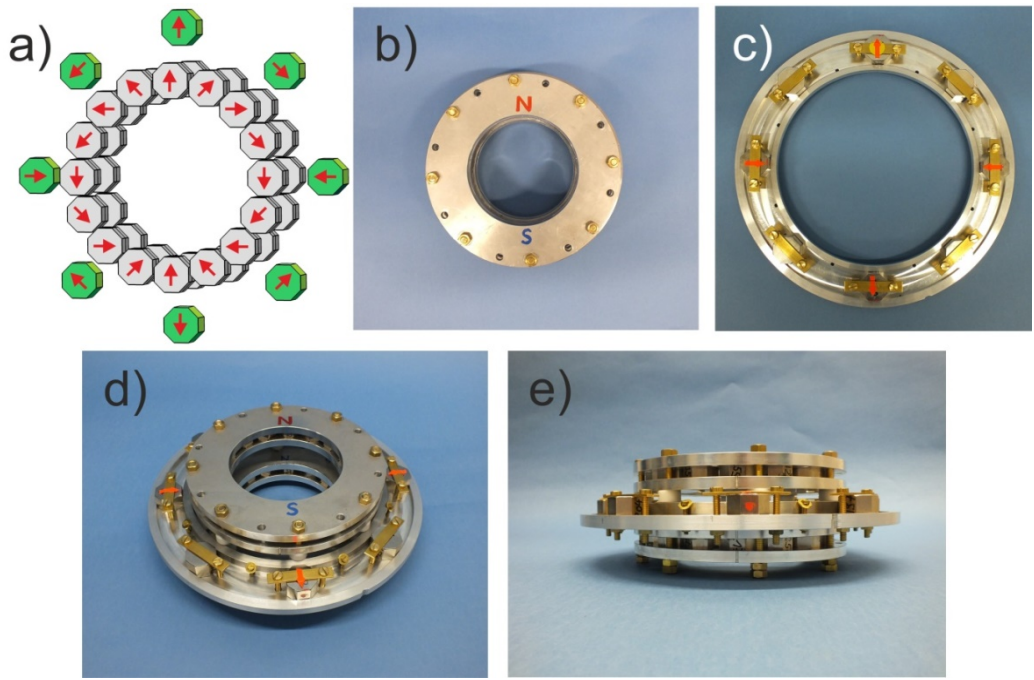

Fig. 4: Demonstration model: a) Arrangement of magnets: The inner pair of rings (gray) with 16 magnets each forms the dipole which is surrounded by 8 magnets (green) for the quadrupole. The red arrows indicate the magnetization direction of the magnets. b) Photograph of the dipole (*N* and *S* marking north and south pole) c) Photograph of the quadrupole (arrows marking the poles) d/e) final assembly, allowing manual rotation of the quadrupole around the dipole without noticeable force.

Figure 5 shows FEM-simulations (COMSOL Multiphysics 5.1) of the magnetic field of the dipole sandwich, the quadrupole and their combination. The magnetic field $\vec{B}(\vec{r})$ inside the dipole was exported as a 1 mm spaced regular grid and imported to MatLab (Mathworks) where the local magnetic field gradient, magnetization and resulting force was calculated. For the calculation of the local magnetization a standard Langevin-function was used ($M(B) = M_{sat}(\coth \xi - 1/\xi)$ with $\xi = \rho V B/(k_B T)$. By way of illustration the parameters were chosen to be typical: $M_{sat}$ = 40 Am$^2$/kg, $\rho$ = 1500 kg/m$^3$, $V$ = 4·10$^{-24}$ m$^3$ (spherical SPP with $r$ = 10 nm) and $k_B$ as the Boltzmann constant. The force was then calculated from eqs. [1] and [3].

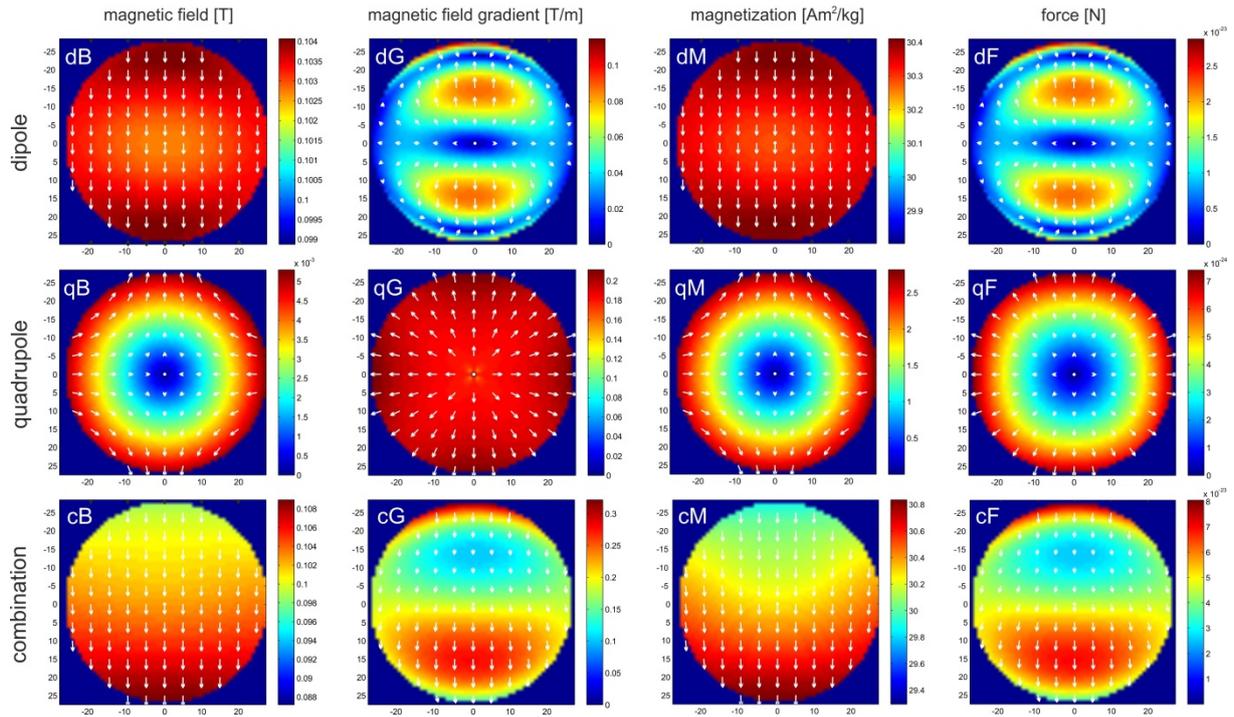

Fig. 5: FEM-simulation of the magnetic field its gradient, the magnetization of a typical SSP and the force on it. Shown is the spatial dependence in the center between the two rings which form the dipole over a circular region of 55 mm in diameter. The colored images in the background show the magnitude of each vector quantity while the white arrows illustrate its directional dependence. Note that the range of each graph was adapted to maximal contrast and is very different as indicated by the color bars right next to each graph. The spatial *x*- and *y*-axes are the same for each graph and given in mm.
The first column to the left shows the magnetic field (second index 'B') as simulated by the FEM-software for the geometry shown in Fig. 4a. From that a numerical derivative (gradient with second index 'G') in both dimensions was calculated in the second column. The magnetization (second index 'M') of a typical SPP was then estimated in the third column (see text for details) and finally in the last column, the force (second index 'F') on such a particle was calculated from the gradient and magnetization using eq. [1] and [3]. This was done for the two sandwiched dipolar Halbach cylinders in the top row (first index 'd') and the quadrupole ring in the middle row (first index 'q') and for their combination in the bottom row (first index 'c') at α = 0.

The field values were also checked experimentally with a Hall-probe (Lakeshore 425 Gaussmeter). Very good agreement was found with the simulated data on the basis that this manual check had an approximate error of 10% on positioning. In Fig. 5 only a small (55 mm in diameter) part of the inner opening is displayed. This was done to clarify the function of dipole, quadrupole and their combination by comparison of $\vec{B}(\vec{r})$, $\vec{\nabla}|\vec{B}(\vec{r})|$, $\vec{M}(\vec{r})$, and $\vec{F}(\vec{r})$ for this prototype. If a larger cross-section would have been chosen (see Fig. 6a) the strong difference in scaling of these vector fields would cover the clarity.

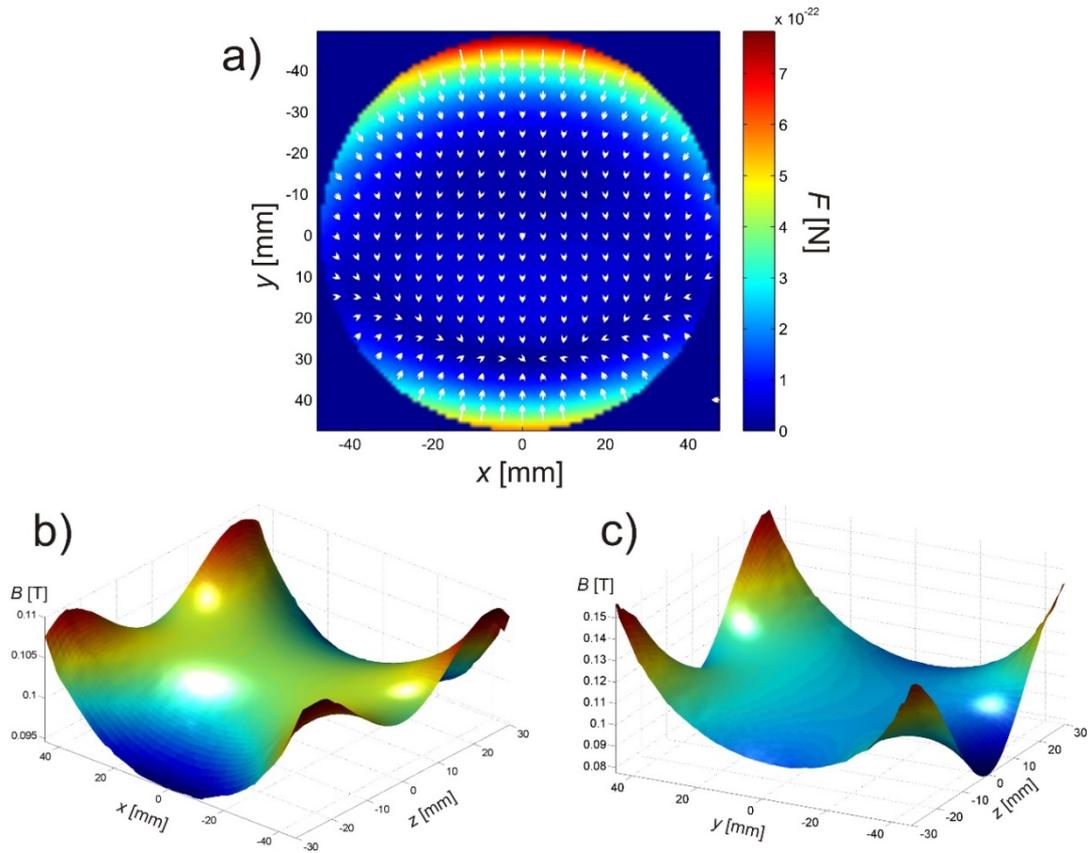

Fig. 6:  a) Same display as in Fig. 5cF for the combined dipol/quadrupole but showing a larger spatial range of 96 mm in diameter (i.e. the inner diameter of the Petri dishes used) b/c) FEM simulation: magnitude of the magnetic field shown as surface plots with additional lighting. b) showing a spatial range of -46 mm ≤ $x$ ≤ 46 mm and -30 mm ≤ $z$ ≤ 30 mm at $y$ = 0. c) Same as b) but for -46 mm ≤ $y$ ≤ +46 mm at $x$ = 0.

These inhomogeneities are caused by the very limited design options of having only enough magnets for two dipole rings. Therefore, the field was preferably optimized in the central region mainly by adjusting the size of the gap between the rings. Nevertheless, Fig. 6a shows an equivalent plot to that of Fig. 5cF for the force over a larger region of 96 mm (i.e. the size of the used samples). The magnitude of the magnetic field along the third $z$-direction is shown as surface plots in Figs. 6b and 6c. From these plots a symmetric saddle shape can be recognized in the $xz$-plane (Fig. 6b) while a similar but flatter and linearly shifted shape is present in the $yz$-plane (Fig. 6c). The linear shift is due to the additional field of the gradient along that direction. This shape is also far from optimal due to the limited length of the dipole. Therefore, sufficient attraction of the SPPs can only be expected in the center of the prototype being in an unstable equilibrium. Whenever the particles will move away from that central region they will be attracted to the maxima at the two dipoles (red peaks in Figs. 6b and 6c). Exactly this behavior was also observed experimentally and hampered a controlled movement in the plane because the particles were drawn out of the interaction plane of the gradient from the quadrupole. In order to work around this problem flat Petri dishes (96 mm in diameter, 20 mm high) were used to confine the particles. Additionally, it was tried to keep the particles floating at the interfacial layer of two immiscible fluids (water and dodecane). Dodecane (Sigma-Aldrich) was chosen due to its relatively low vapor pressure and its dynamic viscosity $\eta = 1.5$ mPas at 25°C being close to that of water ($\eta = 0.9$ mPas at 25°C). Some of the tested SPPs were found floating steadily at the interface for hours.

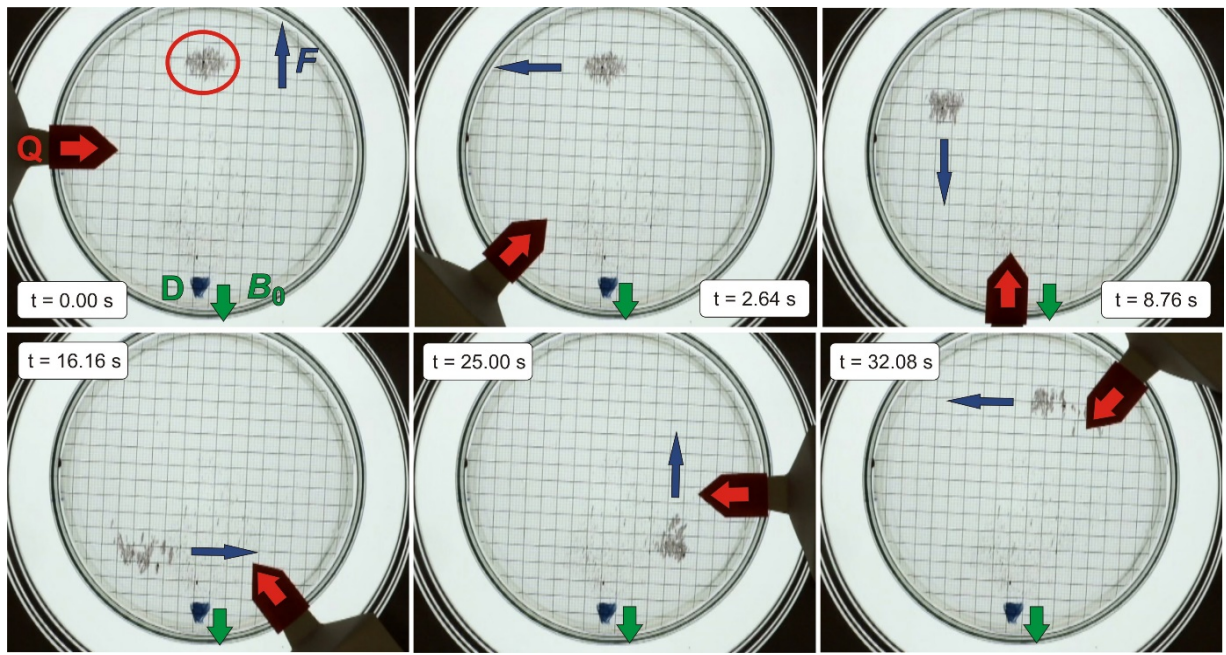

Fig. 7:  Snapshots from a movie of 30 µm particles (Mag 30µm, red encircled in first image) moved inside the demonstration model (Fig. 4) by manual rotation of the Halbach quadrupole: The homogeneous field is along the vertical axis (green arrow). The position of the quadrupole is indicated by a paper arrow (additionally marked by a red arrow). For clarification the resulting force is shown by a slender blue arrow. The SPPs were supported in the third dimension by the interface between a water and dodecane layer inside a Petri dish of 96 mm in diameter. (Full video is available in the supplementing material). At the bottom of the dish a millimeter scale transparency with thicker lines at each 5 mm was attached and the entire setup placed on a photographic light table. The inserts show the time stamp in seconds.

For a demonstration how easily this prototype can move SPPs, spherical magnetite particles of 30 µm in diameter (Mag 30 µm, see supplementing material and Tab. 1 for details) were used. They were carefully placed at the water/dodecane interface of a Petri dish. Figure 7 shows snapshots from a movie (see supplementing material) where the quadrupole was rotated in five intervals of ca. 45° each. This moved the 30 µm particles along a rough square with velocities varying from 3.9 mm/s to 6.4 mm/s.

Similar experiments were carried out with a variety of commercially available hydrophilic SPPs and one sample of lipophilic cobalt ferrite ($CoFe_2O_4$) nanoparticles made by the group of Wolfgang Tremel (inorganic chemistry, University of Mainz) [38]. It was always tried to keep the particles floating at the interface of water/dodecane by carefully inserting them in the interfacial region via a syringe. If this did not work, the particles sank to the bottom of the Petri dish where they seemed to interact with the glass by spreading and forming a layer. They could still be moved by magnetic forces but not as a whole but rather like an unwinding carpet roll (for details see supplementing material).

The velocities were measured optically like in Fig. 7 and compared with theoretical values calculated from eq. [12] to [13]. The essential properties, calculated and measured values are summarized in Tab. 1 (further details can be found in the supplementary part).

Tab. 1: Using eqs. [12] and [13] the theoretical velocity was estimated by $v = 2\rho r^3 M G / (9\eta R)$ for spherical particles of (geometric) radius $r$ and hydrodynamic radius $R$. Their magnetization, $M$, was estimated for a field of 0.1 T. The particles are described in detail in the supplementing material. The $v_W$ is the calculated velocity in water ($\eta$ = 0.9 mPs) and $v_D$ the one in dodecane ($\eta$ = 1.5 mPs) with $G$ = 0.2 T/m (cf. Fig. 7). The asterisk marks experiments where most particles sank to the glass bottom of the Petri-dish. The last row gives the measured velocity, $v_m$, see supplementary material for details on the analysis (videos are available for the last two particles).

| Particle type and diameter | $R$ [nm] | $v_W$ [mm/s] | $v_D$ [mm/s] | $v_m$ [mm/s] |
|---|---|---|---|---|
| SPIO 50 nm | 40 ± 15 | $(15 \pm 5) \cdot 10^{-8}$ | $(9 \pm 3) \cdot 10^{-8}$ | 0.06 ± 0.02* |
| SPIO 130 nm | 65 | $2.8 \cdot 10^{-8}$ | $1.7 \cdot 10^{-8}$ | 0.44 ± 0.02* |
| PSM 3 µm | 1500 | 0.0007 | 0.0004 | 0.1 ± 0.07* |
| Mag 30 µm | 17500 ± 7500 | 0.08 ± 0.07 | 0.05 ± 0.04 | 5.2 ± 0.5 |
| CoFe 75 nm | 37.5 ± 12.5 | $(2.4 \pm 1.0) \cdot 10^{-5}$ | $(1.4 \pm 0.5) \cdot 10^{-5}$ | 14 ± 2 |

## 4. Results and Discussion

Dipole and quadrupole were individually tested for their ability to exert forces on the Mag 30 µm particles. This was done in the same way as shown in Fig. 7. In the dipole alone the particles did not move by themselves from the starting point in the center. However, when mechanically swirled up they slowly wandered towards the poles and concentrated at two spots about 25 – 30 mm away from the center. These spots are also reproduced by the simulations. The graph in Fig. 5dF shows two areas at $y = \pm 23$ mm where the force converges. This effect will be discussed later. The quadrupole alone also showed no force on particles being placed in its center. After swirling them up they accelerated straight in radial directions to the rim (cf. Fig. 5qF), however this process was initially also very slow [39]. The dipole cannot exert a strong force, because the gradients are relatively weak on highly magnetized particles, while the quadrupole possesses an almost constant gradient (cf. Fig. 5qG) but the particles are not magnetized in the center and will experience no strong forces unless being far off the central region.

When both, dipole and quadrupole, are combined the SPPs will be moved in a single direction independent of their starting position and this direction can be simply changed by rotating the quadrupole. This is clearly shown in Fig. 7 and demonstrates that the instrument works as planned and calculated. In this figure it can also be nicely seen how the particles align – like iron filings – along the flux lines of the dipole during the entire movement procedure. They are moved in this alignment by the force of the quadrupole, which acts as predicted at twice its angle. A closer inspection of this movement however shows some irregularities in particle direction, spread and speed. This can be explained from an examination of the magnetic field of the dipole in Fig. 5dB. Clearly the field is not perfectly homogeneous showing a minimum in the center. More interesting is the spatial derivative of this field in Fig. 5dG, where two distinct extrema are observed causing forces (cf. Fig. 5dF) in opposite directions. These local gradients of the dipole are only about a factor of 2-3 smaller than the gradient generated by the quadrupole (cf. Fig. 5qG) and cannot be neglected. As a consequence the force of the combined system in Fig. 5cF has a minimum and a maximum close to the center. Similarly non constant velocities are also observed when moving the particles around experimentally. As shown in detail in the supplementary material the velocities are changing in similar pat-terns as in Fig. 5. However, the effect appears less dramatic than suggested from these simulations. Just like the problem to keep the particles in the center of the third dimension (cf. Fig. 6c/d), this is clearly a consequence of the simple magnet design from limited possibilities. It should be mentioned that Halbach magnets with higher homogeneities and fields have been constructed and are even commercially available with homogeneities of $10^{-8}$ at field strength up to almost 2 T [40].

Nevertheless, the imperfect fields of this prototype have two advantages. Firstly the field turned out to collect the particles at a spot about 3 cm away from the center where the force directions converge (in Fig. 6a towards the bottom, Fig. 7 at 0 s, red encircled). This must be a two dimensional effect and as Fig. 6c shows the $z$-component has a local maximum there. This feature serves nicely for a 2D-application as it collects the particles before reaching the vessel walls. The other advantage is that the concepts of directionality and easy control presented in the theoretical section can be convincingly demonstrated even with such a simple design.

Another surprising observation is that the measured velocities of all particles are two to six orders of magnitude bigger than the calculated ones. None of the experimental parameter in eq. [13] does have an uncertainty that could explain this discrepancy, hence the underlying assumptions must be not correct. As it can be seen from Fig. 7 (and the video of the CoFe particles in the supplementing magterial), the particles form long chains due to their magnetic orientation in the homogeneous field and the resulting mutual attraction. The magnetic moment which interacts with the field gradient is then that of the chain made from a very large number of particles. Closer inspection of the video data also shows a distribution of velocities which clearly scales with the length of the individual chains. Additionally, the effective hydrodynamic radius depends on the direction of the gradient. If this train of particles moves parallel to its long axis, the hydrodynamic radius is close to that of a single particle, while it is the sum of all particles for a movement in perpendicular direction. However, for a quantification of this effect the fields of the apparatus must be at least an order of magnitude more homogeneous and a suitable hydrodynamic model needs to be developed for this particular situation. Nevertheless, this observation might be interesting for MDT because the combination of a homogeneous with a graded field causes this cooperative ("iron filing") effect, which increases the magnetic moment of this super-particle by orders of magnitude while keeping its hydrodynamic radius essentially at the radius of a single particle (for $\vec{G} \parallel \vec{B}_0$).

# 3.  Summary/conclusion

The aim of this work is to propose a new magnetic device for actuating, steering and imaging nanoparticles in MDT or similar applications. This device consists solely of permanent magnets and allows controlling angle and strength of the magnetic force simply by mechanical rotation. If the system is furthermore equipped with an rf-resonator it additionally enables imaging of the particle position via magnetic resonance. The local forces are accessible by simple analytical equations and predicting the path of particles in a non-inhibiting environment is easily possible without additional feedback or control. The principal concept of combining dipolar with quadrupolar magnetic fields was realized with Halbach-cylinders which have no strong stray fields and can easily be rotated in a nested arrangement. Using modern rare-earth permanent magnets, field strength up to 2 T and gradient strength in the order of 10 or 20 T/m are realistic with this design. However, this design is not invariant to scaling as shown in part 2.5, and gradient strength must be traded for angular precision in certain locations when the sampled region is increased. However, strategies were presented how this problem can be reduced, e.g. by synchronous rotation of dipole and one quadrupole. Other combinations of dipolar and quadrupolar rings might also be useful depending on the application and the complexity of operation. For instance, bringing magnetic particles inside a biological system, locate and hold them at a specific initial spot will become complicated in the presence of strong magnetic fields and gradients. Even if the radial gradients are removed ($\beta = 90°$), the axial gradient will remain and a patient or syringe containing the magnetic particles will have to cross it on its way to the application spot. The same holds once the particles have reached their destination and shall not exert additional forces thereafter. If this turns out to be a major concern, all field (homogeneous and gradients) can be removed by yet adding another magnetic dipole ring to make $B_0$ disappear as suggested in [24].

For these reasons the concept of combining and rotating dipolar and quadrupolar fields was discussed in greater detail, because this allows the design of magnetic systems which are tailored to their particular task. Therefore, a demonstration of this concept by a machine that could do the everything in one was waived in favor of a proof of principle by a simple prototype. With this simple instrument larger, still visible particles could be moved at almost constant speed over a large fraction of the inner opening of the magnet. The direction of movement was simply adjusted by rotating the outer quadrupole by hand. Some additional features (scaling of field [24] and imaging [34]) of a more complex magnet system with two quadru-poles could not be tested with this simple prototype but were already previously demonstrated.

Of course, it is possible to generate stronger gradients and forces by resistive [41] or superconducting magnets [42], but typically in a static and/or single-sided fashion. Using permanent magnets instead has the advantage that no cooling or charging of the coils is necessary, making this instrument maintenance-free, silent and almost non-magnetic on the outside (no radial stray-fields). This is possible because new rare-earth permanent magnetic materials (e.g. NdFeB) create extremely strong magnetic fields ("*a current loop of 1000 A can be replaced by an approximately 1-mm-thick NdFeB permanent magnet block*" [43]), and although the gradients might be weaker, they seem to be sufficient and their application is not time-sensitive. These gradients will probably also do not cause a safety issue, although they are very strong, because there is no need to change their strength or direction at such a speed that the currents induced in the tissues will exceed critical limits. Their sheer strength is not bigger than that at fringe of an MRI superconducting magnet.

An interesting and unintentional observation is the enormous increase of particle speeds in the prototype. This effect was explained by the formation of very long particle chains due to the strong orientation and magnetization in the homogeneous field causing mutual attractions. The magnetic moment of such a chain will then be the sum of all particles while the hydrodynamic resistance may stay close to that of a single particle. This effect needs further study in well-defined magnetic fields and a better suited hydrodynamic model to describe it. However, this effect might explain why even the simple prototype worked surprisingly well.

Taken together, the authors hope that the presented ideas can be used to construct magnetic devices which will meet the first three challenges quoted in the introduction; i.e. provide sufficiently strong forces deep in the body and allow real-time imaging. In view of eq. [10] there is probably no need for sophisticated control mechanisms. However, understanding the particle velocities and motion under real conditions still need investigations which will be studied next with an improved setup. Hopefully, this work will trigger new technical approaches to improve MDT applications.

# Acknowledgements:

First of all PB wants to thank Matthias Barz (Institute of Organic Chemistry, Uni Mainz) for triggering the idea which stimulated this work. He also wants to thank his wife, Friederike Schmid (Institute of Physics, Uni Mainz), for checking the manuscript and cleaning up the math. The authors are thankful to Martin Klünker (Institute of Inorganic Chemistry, Uni Mainz) for providing the CoFe particles. The DFG funded Collaborative Research Center *"Nanodimensional polymer therapeutics for tumor therapy"* – SFB 1066 (in particular Rudolf Zentel) has to be accredited for financial support and last but not least Werner Heil (Institute of Physics, Uni Mainz) is to acknowledge for tolerating and supporting this side activity in his research group.

# Supplementary Material to

## "Permanent Magnet System to guide superparamagnetic particles"

Olga Baun and Peter Blümler*

Institute of Physics, University of Mainz, Germany

### S1. Properties of used superparamagnetic particles:

All hydrophilic particles (SPIO, PSM and Mag) were obtained from micromod GmbH, Rostock, Germany. Their properties are summarized below with the abbreviations used in Tab. 1:

<u>SPIO 50 nm</u>: nanomag-D-spio (product number 79-00-501) dextrane-ironoxide-composite with a plain surface (OH groups), $M(0.1T) = 56$ Am$^2$/(kg iron) with an iron mass concentration of 0.096 per particle mass resulting in a (mass) magnetization of $M(0.1T) = 5.4$ Am$^2$/(kg particles) and a density $\rho = 1400$ kg/m$^3$. They are synthesized by precipitation of iron oxide in the presence of dextrane forming a matrix (40 kDa) with an iron concentration of 2.4 mg/ml. The hydrodynamic radius can vary between 50 nm and 110 nm. Therefore, for the calculation a value of 80 nm ± 30 nm was used.

<u>SPIO 130 nm</u>: nanomag-D (product number 09-00-132) dextrane-ironoxide-composite with a plain surface (OH groups), with a (mass) magnetization of $M(0.1T) = 5.3$ Am$^2$/(kg particles) and a density $\rho = 2500$ kg/m$^3$. They are synthesized via the core-shell method with a 75-80 w/w magnetite core and a dextrane shell (40 kDa).

<u>PSM 3 μm</u>: micromer-M (product number 08-01-303) magnetic polystyrene matrix particles with a NH$_2$ modified surface. They are monodisperse particles which consist of magnetite around an organic matrix of a styrene-maleic acid-copolymer. The (mass) magnetization is $M(0.1T) = 5.4$ Am$^2$/(kg particles) with a density $\rho = 1100$ kg/m$^3$.

Mag 30 µm:   PLA-M (product number 12-00-304) magnetite (40% w/w) in a matrix of poly(D,L-lactic acid) with a molecular weight of 17 kD and a plain surface, the size range is 20-50 µm. Therefore, a geometric = hydrodynamic radius of 35 µm ± 15 µm was used. The (mass) magnetization is $M(0.1T) = 4.3$ Am$^2$/(kg particles) with a density $\rho = 1300$ kg/m$^3$.

CoFe 75 nm: Lipophilic cobalt ferrite ($CoO \cdot Fe_2O_3$) particles 50-100 nm in diameter). They were synthesized and provided by the group of W. Tremel [38]. $M(0.1T) = 65$ Am$^2$/(kg particles) and $\rho \approx 5300$ kg/m$^3$.

## S2.   Analysis of particle velocity

The velocity of all particles was determined by the setup described in the main publication. However, their very different motion requested different analytical strategies which are described in the following:

SPIO 50 nm:

When a drop of the solution was injected at the interface between water and dodecane, the particles immediately sank to the bottom of the dish where they spread on (cf. Fig. S1a) forming a layer which can be approximated by an ellipse. This ellipse changes its center of mass (origin) and area under the application of magnetic force (cf. Fig. S1b). For an estimation of an average velocity the motion of the center of the ellipses was chosen and the error was estimated from the minimum velocity at its upper edge.

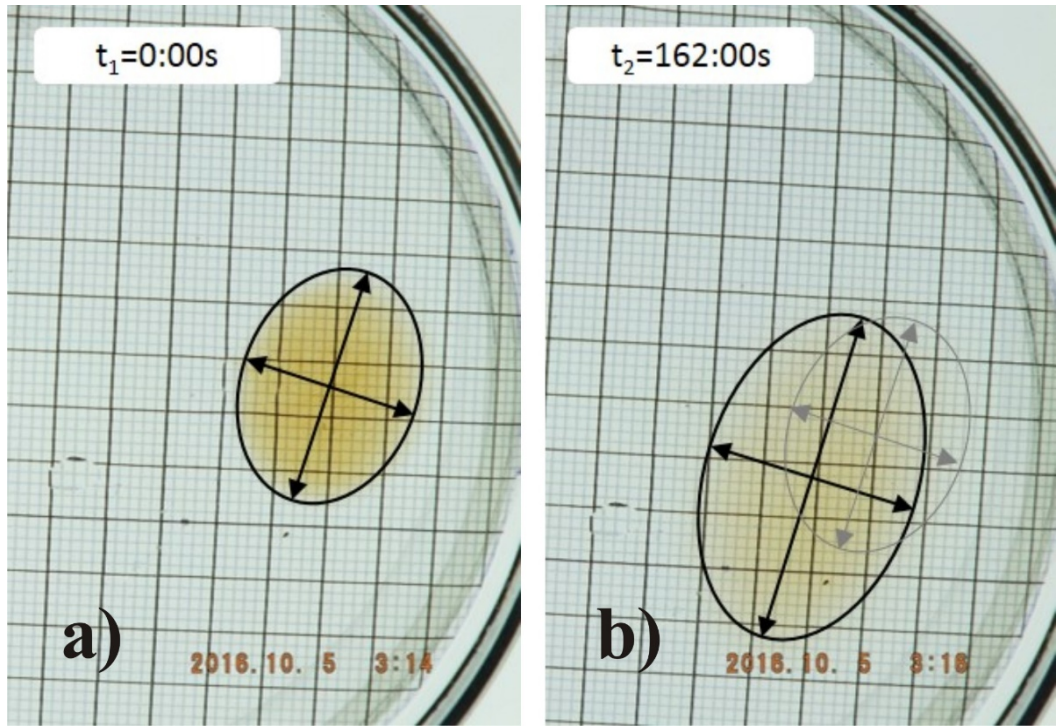

Fig. S1: Photographs of SPIO 50 nm particles at time a) $t_1$=0:00s and b) $t_2$=162:00s. At both times, the particle drop was marked by an ellipse including axes. The intersection of the axes corresponds to the center of mass, which moves in time in the direction of the applied force, while the particles at the upper edge change their position only slightly.

SPIO 130 nm:

In this experiment most of the particles also sank to the bottom of the dish, where they seemed to interact with the glass surface (Fig. S2a). When the magnetic force was applied it looked as if a layer on top of the bottom layer started to move (like an unwinding carpet roll) uniformly with a distinct front. This front could be marked in four consecutive snapshots and with the assumption of a linear propagation, the velocity of several points on each front were calculated. From these four values the mean and the standard deviation were determined.

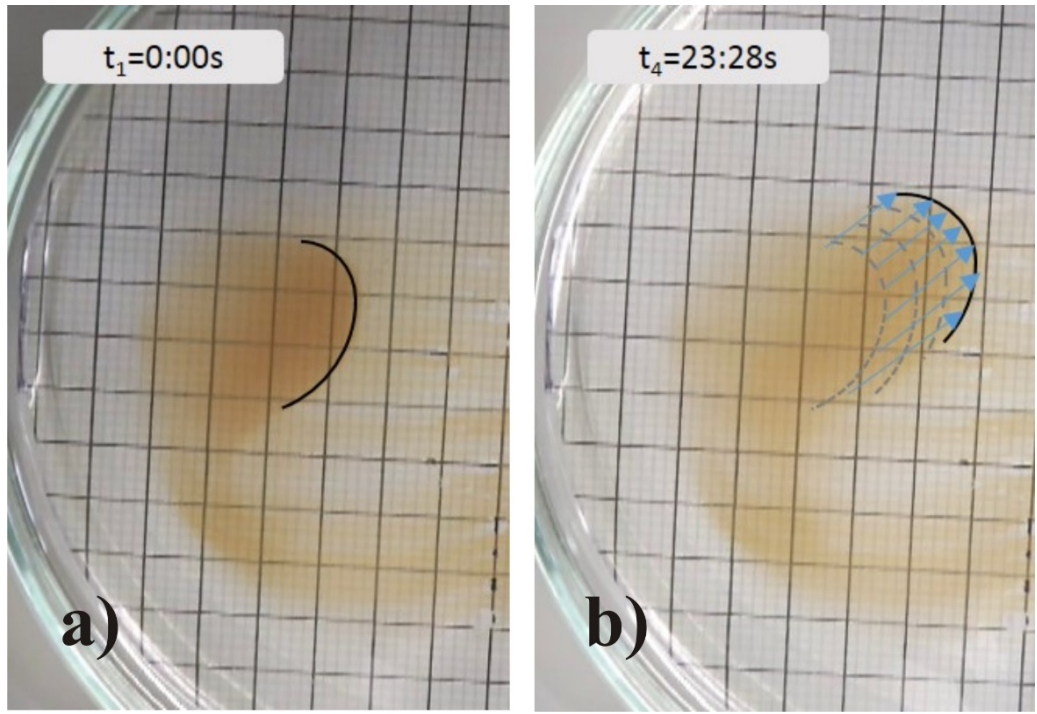

Fig. S2: Photographs of SPIO 130 nm particles at a) $t_1$=0:00s and b) $t_4$=23:28s. a) The marking of the front with the maximal accumulation of particles is indicated. b) Movement of the marking at time $t_4$ (solid line). The three dashed lines show the position of the particle front at times $t_1$, $t_2$=6:44s, and $t_3$=14:68s.

PSM 3μm:

When these particles were injected to the liquid, they quickly fanned out in the direction of the magnetic force, but also rapidly sank to the bottom of the Petri dish. Once they reached the bottom they only showed a comparatively small tendency to move. Figure S3a is a sor of projection of the initial motion of the particles while sinking through the liquid. Although the shape of this U-shaped figure changed slowly over time, it was very difficult to identify reliable features. Nevertheless some fronts were marked in Fig. S3 by boxes and horizontal extensions used to calculate a mean velocity and its standard deviation. However, this value has a great systematic error.

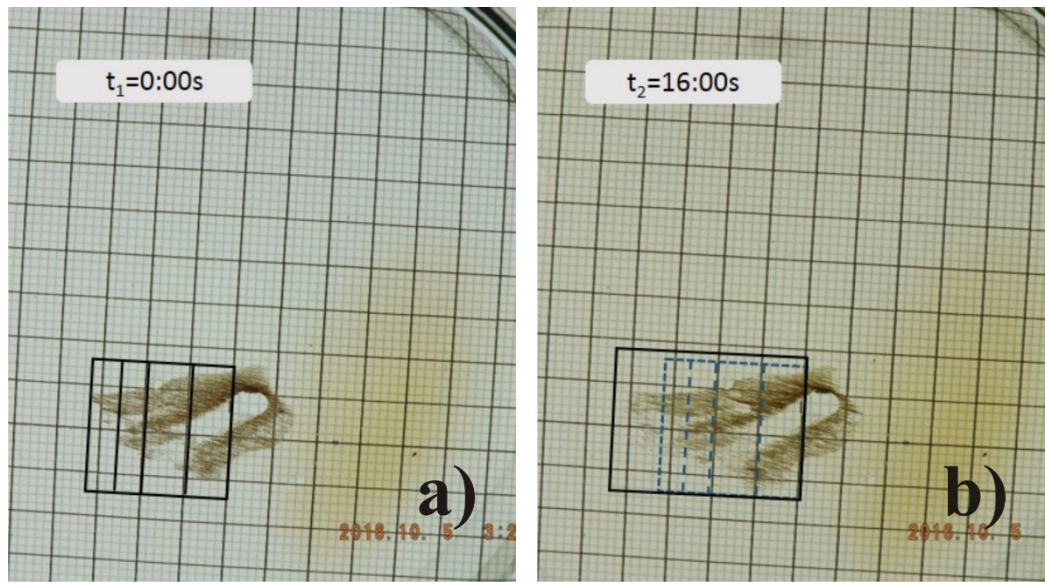

Fig. S3: Photographs of PSM 3µm particles at the times a) $t_1$=0:00s and b) $t_2$=16:00s. A movement of the particles can be clearly recognized, but the assignment of the particles is difficult. Particles in in four marked ranges were optically chosen and used for an estimation of velocities.

Mag 30 µm:

These particles liked to stay at the interface between dodecane and water. Their movements was also fast enough to take a video (video1.avi). From various snapshots (see also Fig. 7 of main article) individual clots of particles could clearly be tracked (as indicated by colors in Fig. S4). Their displacements were determined in 8 pictures, from which a mean velocity and its standard deviation were calculated.

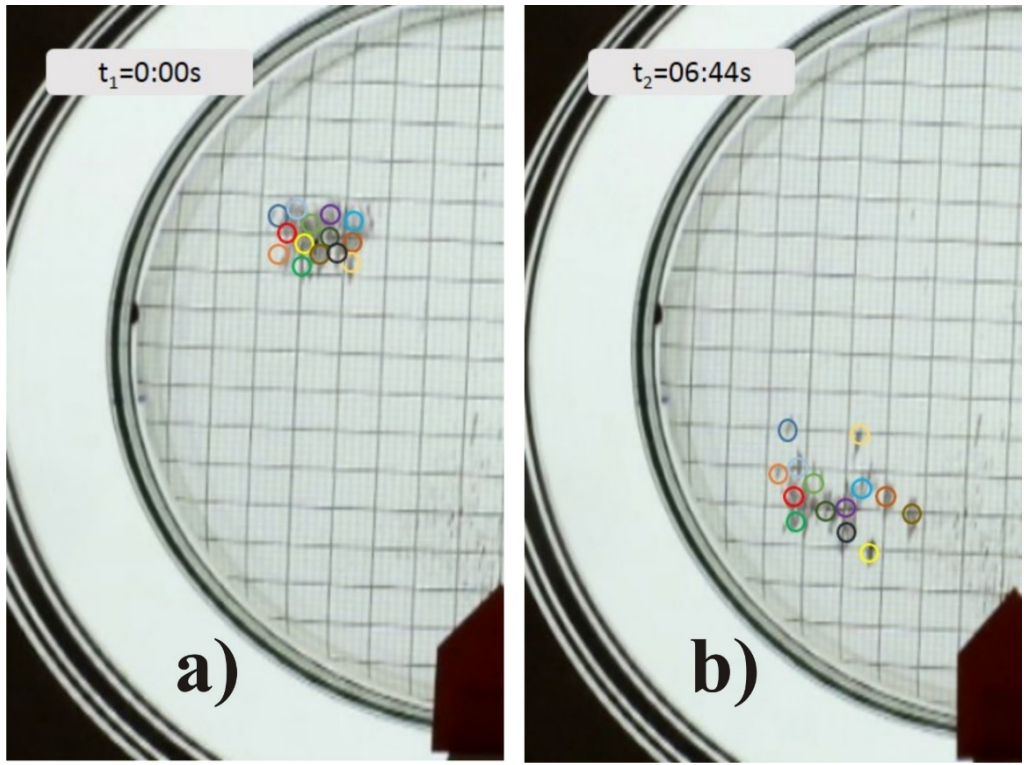

Fig. S4: Snapshots of the Mag 30 µm particles from video1. The particle drop was divided into 14 areas, which were tracked using the movie. By assigning the particles to the different times, a mean value and an error of the velocity could be determined.

CoFe 75 nm:

Because these particles have a tendency to form agglomerates, they were treated for 10 min in an ultrasonic bath before the experiment. Their lipophilic coating allowed very easy application, because they dissolved in the top dodecane layer and were pulled down towards the interface by gravity and magnetic force were they stayed. Like the Mag 30 µm particles they showed speeds that allowed to record a video (video2.avi). In difference to the Mag 30 µm particles they showed a more pronounced tendency to agglomerate to very long chains with correspondingly higher velocities (probably due to their higher magnetic moment). This can be appreciated because the appearance of the particles is similar to that of the 30 µm particles before, although the individual particles should not be visible. Otherwise the analysis was identical to that of the Mag 30 µm particles.

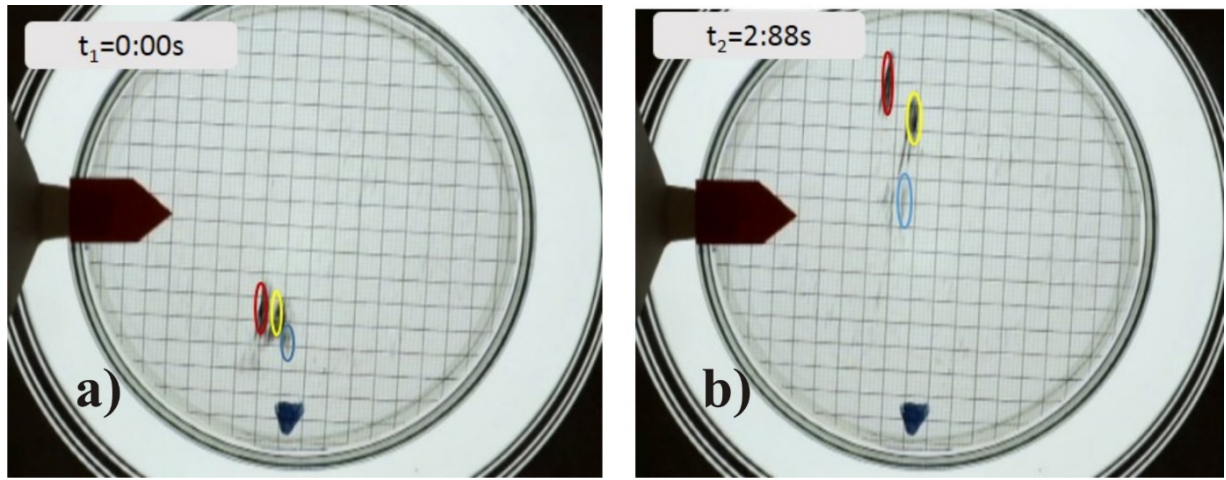

Fig. S5: Snapshots of the CoFe 75 nm particles from video2 at times a) $t_1$=0:00s and b) $t_2$=2.88s. In these two images the strong tendency of these particles to form large chains can clearly be seen. The speed of the chains seems to have a good correlation to their length. The individual chains were approximated by ellipses. The displacement of the centers of these ellipses was used to calculate the individual velocities.

## S3. Videos

The videos are acquired with a regular digital camera (Fujifilm Finepix HS30) using the setup described in the article.

**video1.avi**: A 42 s long real time sequence (25 frames/s) showing the Mag 30 µm particles being moved roughly along a square. (data length 151 MByte)

**video2.avi**: A 39 s long real time sequence (25 frames/s) showing the CoFe 75 nm particles being moved roughly along a square. (data length 141 MByte)

→ Both videos can be uploaded from
https://fileshare.zdv.uni-mainz.de/rKD2FwqBnk7bM09mdtB3OQ.repo
(use anonymous login)